\documentclass[]{spie}  

\usepackage{amsmath,amsfonts,amssymb}
\usepackage{graphicx}
\usepackage[acronym]{glossaries}
\usepackage{xcolor}
\usepackage{marginnote}

\usepackage{acronym}
\newacro{ccd}[CCD]{Charged Coupled Device}
\newacro{lsstcam}[LSSTCam]{Legacy Survey of Space and Time Camera}
\newacro{ToO}[ToO]{Target-of-Opportunity}
\newacro{BH}[BH]{black hole}
\newacro{BBH}[BBH]{binary black hole}
\newacro{BNS}[BNS]{binary neutron star}
\newacro{GCN}[GCN]{General Coordinates Network}
\newacro{NSBH}[NSBH]{neutron star black hole}
\newacro{lsstcomcam}[lsstcomcam]{LSST Commissioning Camera}
\newacro{OCS}[OCS]{Observatory Control System}
\newacro{LOVE}[LOVE]{LSST Operations and Visualization Environment}
\newacro{dL}[$d_L$]{luminosity distance}
\newacro{GW}[GW]{Gravitational-wave}
\newacro{FES}[FES]{Filter Exchange System}
\newacro{GRB}[GRB]{gamma ray burst}
\newacro{NS}[NS]{neutron star}
\newacro{DIA}[DIA]{difference-imaging analysis}
\newacro{DES}[DES]{Dark Energy Survey}
\newacro{CBC}[CBC]{compact binary coalescence}
\newacro{SNR}[SNR]{signal-to-noise ratio}
\newacro{LVK}[LVK]{LIGO-Virgo-KAGRA}
\newacro{EM}[EM]{electromagnetic}
\newacro{LSST}[LSST]{Legacy Survey of Space and Time}
\newacro{WFD}[WFD]{Wide Fast Deep}
\newacro{BTS}[BTS]{Base Test Stand}
\newacro{CTIO}[CTIO]{Cerro Tololo Inter-American Observatory}
\newacro{DECam}[DECam]{the Dark Energy Camera}
\newacro{DDFs}[DDFs]{Deep Drilling Fields}
\newacro{SSM}[SSM]{sub-solar mass}
\newacro{DDF}[DDF]{Deep Drilling Field}
\newacro{FBS}[FBS]{Feature Based Scheduler}
\newacro{SCOC}[SCOC]{Survey Cadence Optimization Committee}
\newacro{HEv}[$\text{HE-}\nu$]{high-energy neutrinos}
\newacro{PHAs}[PHAs]{potentially hazardous asteroids}
\newacro{GalSN}[Gal-SN]{galactic supernovae}
\newacro{Super-K}[Super-K]{Super-Kamiokande}
\newacro{SV}[SV]{science validation}
\newacro{FOV}[FOV]{field-of-view}
\newacro{JPL}[JPL]{Jet Propulsion Laboratory}
\newacro{usdf}[USDF]{United States Data Facility}
\newacro{SCiMMA}[SCiMMA]{Scalable Cyberinfrastructure to support Multi-Messenger Astrophysics}
\newacro{SNEWS}[SNEWS]{Supernova Early Warning System}
\newacro{Super-K}[Super-K]{Super-Kamiokande}
\newacro{EFD}[EFD]{Engineering Facility Database}
\newacro{SAL}[SAL]{system abstraction layers}
\newacro{CSC}[CSC]{Commandable \ac{SAL} Component}
\newacro{PPDB}[PPDB]{Prompt Products Database}
\newacro{rsp}[RSP]{Rubin Science Platform}
\newacro{O4}[O4]{Observing Run 4}
\newacro{RTO}[RTO]{Rubin ToO Observer}
\newacro{TNS}[TNS]{Transient Name Server}
\newacro{IR1}[IR1]{Intermediate Run 1}

\usepackage[colorlinks=true, allcolors=blue]{hyperref}

\title{The Rubin Observatory Target-of-Opportunity System in the First Year of Operations}

\author[a]{Sean Patrick MacBride}
\author[b]{R. Lynne Jones}
\author[b]{Peter Yoachim}
\author[c]{Tiago Ribeiro}
\author[d]{Leanne P. Guy}
\author[e,f]{Shreya Anand}
\author[b]{Erin Leigh Howard}
\author[b]{Ian S. Sullivan}
\author[g]{Daniel L. Wang}
\author[b]{Eric C. Bellm}
\author[f]{Robert Armstrong}
\author[i]{W. M. Wood-Vasey}
\author[j]{Alex Drlica-Wagner}
\author[j]{Kenneth Herner}
\author[k]{Gautham Narayan}
\author[l]{Tatiana Acero-Cuellar}
\author[l,m,n]{Federica Bettina Bianco}
\author[o,p,q]{Igor Andreoni}
\author[r]{Bruno O. S\'anchez}
\author[s]{John Banovetz}
\author[c]{Anastasia Alexov}
\author[c]{Erik Dennihy}
\author[c]{Robert D. Blum}
\author[t]{Yousuke Utsumi}
\author[a]{Marcelle Soares-Santos}
\author[h]{Robert H. Lupton}
\author[c,u]{Keith Bechtol}
\author[c,b]{\v{Z}eljko Ivezi\'{c}}
\author[d]{Kris Mortensen}
\author[d]{Paulina Venegas}
\author[e,d]{Yijung Kang}
\author[d]{Alysha B. Shugart}
\author[d,g]{Narayan Khadka}
\author[d]{Kshitija Kelkar}
\author[d]{Danica \v{Z}ilkov\'a}
\author[g]{Kevin Fanning}
\author[v]{Christopher W. Walter}
\author[w]{Chris Weaver}
\author[k]{Fabian Araneda-Baltierra}
\author[x]{Andreja Gomboc}
\author[y]{J. Anthony Tyson}
\author[b]{Colin Orion Chandler}
\author[z]{Pedro H. Bernardinelli}
\author[b]{Devanshi Singh}

\affil[a]{Physik-Institut, University of Zurich, Winterthurerstrasse 190, 8057 Zurich, Switzerland}
\affil[b]{University of Washington, Dept.\ of Astronomy, Box 351580, Seattle, WA 98195, USA}
\affil[c]{NSF-DOE Vera C.\ Rubin Observatory / NSF NOIRLab, 950 N.\ Cherry Ave., Tucson, AZ  85719, USA}
\affil[d]{NSF-DOE Vera C.\ Rubin Observatory / NSF NOIRLab, Casilla 603, La Serena, Chile}
\affil[e]{Kavli Institute for Particle Astrophysics and Cosmology, Stanford University, 452 Lomita Mall, Stanford, CA 94305, USA}
\affil[f]{Department of Physics, Stanford University, 382 Via Pueblo Mall, Stanford, CA 94305, USA}
\affil[g]{SLAC National Accelerator Laboratory, 2575 Sand Hill Rd., Menlo Park, CA 94025, USA}
\affil[h]{Department of Astrophysical Sciences, Princeton University, Princeton, NJ 08544, USA}
\affil[i]{Department of Physics and Astronomy, University of Pittsburgh, 3941 O'Hara Street, Pittsburgh, PA 15260, USA}
\affil[j]{Fermi National Accelerator Laboratory, P. O. Box 500, Batavia, IL 60510, USA}
\affil[k]{University of Illinois, Physics and Astronomy Departments, 1110 W.\ Green St., Urbana, IL  61801, USA}
\affil[l]{Department of Physics and Astronomy, University of Delaware, Newark, DE 19716-2570, USA}
\affil[m]{Data Science Institute, University of Delaware, Newark, DE 19717 USA}
\affil[n]{Joseph R.\ Biden, Jr., School of Public Policy and Administration, University of Delaware, Newark, DE 19717 USA}
\affil[o]{Joint Space-Science Institute, University of Maryland, College Park, MD 20742, USA}
\affil[p]{Department of Astronomy, University of Maryland, College Park, MD 20742, USA}
\affil[q]{Astrophysics Science Division, NASA Goddard Space Flight Center, Mail Code 661, Greenbelt, MD 20771, US}
\affil[r]{Aix Marseille Universit\'{e}, CNRS/IN2P3, CPPM, 163 avenue de Luminy, F-13288 Marseille, France}
\affil[s]{Brookhaven National Laboratory, Upton, NY 11973, USA}
\affil[t]{National Astronomical Observatory of Japan, Chile Observatory, Los Abedules 3085, Vitacura, Santiago, Chile}
\affil[u]{Department of Physics, University of Wisconsin-Madison, Madison, WI 53706, USA}
\affil[v]{Department of Physics, Duke University, Durham, NC 27708, USA}
\affil[w]{Institute for Cyber-Enabled Research, Michigan State University, East Lansing, MI 48824, USA}
\affil[x]{Center for Astrophysics and Cosmology, University of Nova Gorica, Vipavska 13 5000 Nova Gorica, Slovenia}
\affil[y]{Physics Department, University of California, One Shields Avenue, Davis, CA 95616, USA}
\affil[z]{Institute for Data-intensive Research in Astrophysics and Cosmology, University of Washington, 3910 15th Avenue NE, Seattle, WA 98195, USA}

\authorinfo{Author information: E-mail \href{mailto:sean.macbride@physik.uzh.ch}{sean.macbride@physik.uzh.ch}}

\begin{document} 
\maketitle

\begin{abstract}
The NSF/DOE Vera C. Rubin Observatory is a discovery machine, with unprecedented survey speed, which can be used to identify exotic astrophysical transients. In its prime mission, the ten year Legacy Survey of Space and Time will use 3\% of its total time for Target of Opportunity observations, which includes response to gravitational wave events, high energy neutrinos, potentially-hazardous asteroids, and other astrophysical phenomena. Target of Opportunity observations exist outside of the usual LSST operational mode, requiring special attention to maximize performance. We review the Rubin Target of Opportunity system during its first year of Rubin Observatory operations, the Targets of Opportunity pursued since LSST first light, and the overall efficiency of the system.
\end{abstract}


\section{INTRODUCTION}\label{sec:intro}  
The NSF/DOE Rubin Observatory is a discovery machine that will revolutionize astronomy through its unprecedented observing speed and sensitivity. This speed and sensitivity are enabled by unique technical capabilities of Rubin Observatory - the 8.4-meter combined primary/tertiary mirror enables the primary instrument of Rubin Observatory, \ac{lsstcam}, to image a $9.6\deg^2$ \ac{FOV} in one of the six optical filters across the visible spectrum \cite{ScienceDrivers}. A standard exposure of thirty seconds can achieve a 5-$\sigma$ apparent magnitude sensitivity of $\sim24$ in the $r$-band, rapidly monitoring the southern sky for faint objects. The combination of the large {FOV}, deep observations, and fast slew-point-image cycle, makes Rubin Observatory an ideal facility to identify faint astrophysical phenomena.

Rubin Observatory will execute a ten-year survey starting in 2026, named the \ac{LSST}. The observatory’s design, and the implementation of the \ac{LSST}, is driven by four key science themes: probing dark energy and dark matter; taking an inventory of the solar system; exploring the transient and variable optical sky; and mapping the Milky Way. 

The \ac{LSST} is composed of different surveys to achieve these scientific goals, with each survey using a different observing plan\cite{PSTN-056}.

\begin{itemize}
    \item \textbf{\ac{WFD}.} This is the bulk of the \ac{LSST} survey, and is designed to achieve the core science goals of the \ac{LSST}. The \ac{WFD} name is indicative of the observing behavior - observing a \textit{wide} area of the sky ($\sim19.6\text{k} \deg^2$), at a \textit{fast} cadence (each pointing receives a return visit every 2-4 nights), with \textit{deep} observations (single visit depth of $m_r\sim24$). The \ac{WFD} uses slightly more than 80\% of the total survey time.
    \item \textbf{Mini and Micro Surveys}. These are small sub-surveys that cover specific regions of the sky (such as the North Ecliptic Spur, Dusty Plane, or South Celestial Pole) to accomplish specific science goals, often operating with a different cadence from the \ac{WFD}. Minisurveys could use between 3-10\% of the total survey time, while microsurveys could use 1-3\% of the total survey time.
    \item \textbf{\ac{DDFs}.} These are single \ac{lsstcam} pointings (9.6 $\deg^2$ each), which receive intensive observations on a regular basis. Each \ac{DDF} will reach at least 1 magnitude deeper than the \ac{WFD} coadded depths. The \ac{DDFs} use around 7\% of the total survey time.
    \item \textbf{\ac{ToO} observations.} These are observations that are made in response to multimessenger alerts from external physics experiments. This includes optical followup of \ac{GW} events from \ac{LVK} alerts, observations of \ac{PHAs} from JPL-Scout, observations of \ac{HEv} regions from IceCube alerts, and capturing the shock-breakout of a \ac{GalSN} using early identification from \ac{Super-K} experiment. \ac{ToO} observations will use 3\% of total survey time.
\end{itemize}

The \ac{ToO} survey uses a different observing mode of the \ac{FBS} than all other surveys of the \ac{LSST}, requiring a rapid response to time-sensitive astrophysical phenomena and interrupting scheduled observations from other surveys. The \ac{FBS} is an machine-learning enabled scheduling algorithm that executes all of the \ac{LSST} surveys based on weather conditions, science priorities, and observatory performance (see Section \ref{subsec:Scheduler} for further details, or Jones\cite{LynnesPaper}). The time to respond to alert receipt can vary, ranging from mere minutes to hours or even days. Each \ac{ToO} target has a different observing strategy based on observing conditions, target properties, and other astrophysical parameters. The \ac{ToO} observing strategies are the product of community input and were most recently revised in 2024 \cite{RubinToO2024}. These recommendations were incorporated into the survey strategy recommendation issued by the \ac{SCOC}\footnote{The SCOC is a standing committee of Rubin Observatory making survey strategy recommendations to the Observatory Director, responsible for optimizing the survey strategy to maximize the scientific throughput of LSST.} in January 2025 \cite{PSTN-056}.

Rubin is uniquely suited for follow-up of external transient alerts due to its wide field of view, deep imaging sensitivity, and rapid, automated response capabilities. This makes it especially powerful for localizing and characterizing multi-messenger events with large initial positional uncertainties, such as gravitational-wave and high-energy neutrino alerts. Additionally, the rapid response capabilities are power for constraining the orbits of potentially hazardous asteroids, and observing the evolution of a galactic supernovae.

The current implementation of the Rubin \ac{ToO} program considers alerts from the following astrophysical messengers and associated experiments for \ac{ToO} observation:

\begin{itemize}
    \item \textbf{Gravitational waves:} \ac{LVK} Collaboration. \cite{aLIGO}
    \item \textbf{High energy neutrinos:} IceCube Neutrino Observatory. \cite{icecube_whitePaper}
    \item \textbf{Potentially hazardous asteroids:} The NASA \ac{JPL} Scout and \ac{JPL} Sentry systems. \cite{JPLSentry, neoSurveyor}
    \item \textbf{Galactic supernovae:} The \ac{SNEWS} and the \ac{Super-K} experiment. \cite{SNEWS_2_0, superk}
\end{itemize}

Rubin \ac{ToO} alert classes (GW, high energy neutrinos, etc.) have been curated for their potential for scientific discovery using the unique capabilities of Rubin Observatory. The list of alert types and their associated experiments is by no means static, and is reviewed regularly in consultation between the Rubin science community and the \ac{SCOC}.\footnote{Alert classes shown here are accurate as of \today}

\ac{ToO} observations are time sensitive, and valuable enough that we choose to interrupt \ac{LSST} observations to observe these priority fields. The unique nature of \ac{ToO} observations requires different workflows, communication channels, and operations to ensure that \ac{ToO} observations are valuable to the \ac{LSST} and the scientific community. Additionally, the serendipitous nature of \ac{ToO}s means that they can occur at any time during the \ac{LSST}, so readiness, uptime, and low latency is crucial to the success of \ac{ToO} observations.

In this paper, we describe the implementation, workflow, and performance of the Rubin Observatory's \ac{ToO} system during the first year of Rubin Observatory operations. In section \ref{sec:technical_overview}, we provide a technical overview of the \ac{ToO} system. In section \ref{sec:system_integration}, we discuss the integration of the \ac{ToO} system to Rubin Observatory including verification of different Rubin components using simulated \ac{ToO} responses. In section \ref{sec:commissioning_performance}, we discuss the responses to real alerts during the early operations period and efficiency of the \ac{ToO} system. Finally, we discuss the lessons learned from early integration and review the outlook of the \ac{ToO} program for the ten year \ac{LSST} in sections \ref{sec:lessons_learned}-\ref{sec:Conclusion}.
\section{Technical Overview of the Rubin ToO System}\label{sec:technical_overview}
\begin{figure}[b]
    \centering
    \includegraphics[width=\linewidth]{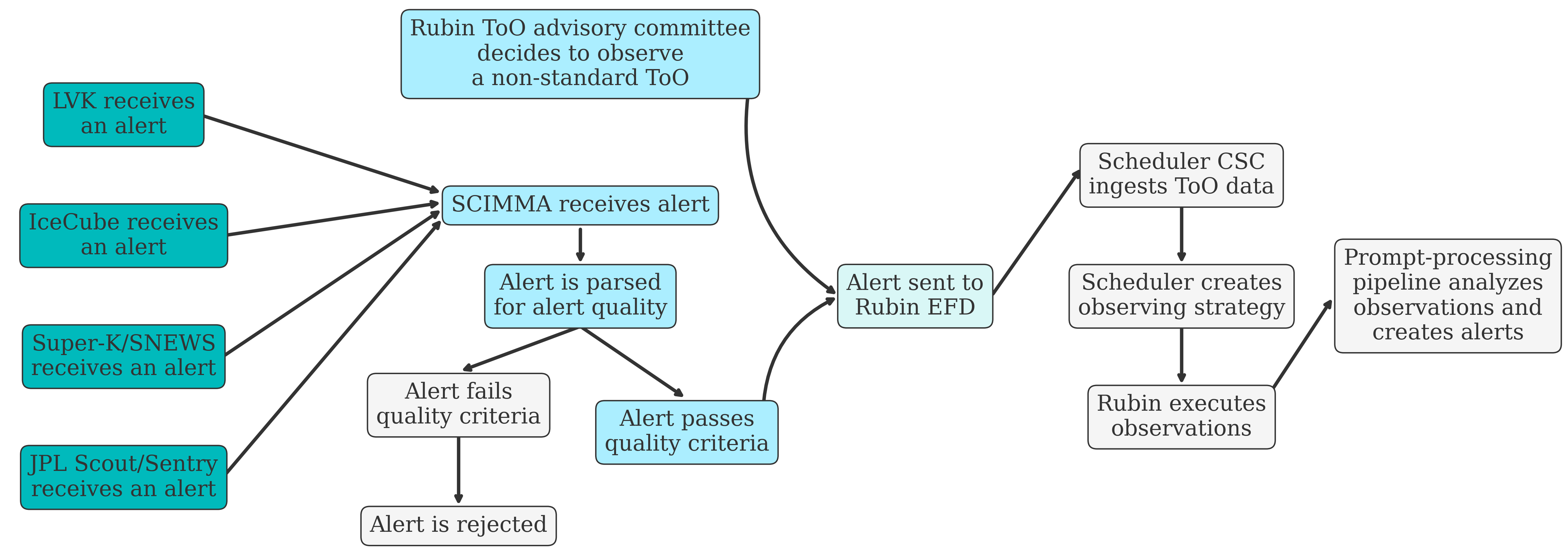}
    \caption{The Rubin \ac{ToO} response workflow. Time flows from left to right.}
    \label{fig:workflowDiagram}
\end{figure}

The Rubin observatory \ac{ToO} system is composed of distinct software components that are responsible for multi-messenger alert receipt from external experiments, updating the Rubin Observatory observing plan, executing observations in the region of interest of a \ac{ToO}, and rapidly processing images to detect variable sources. The high-level overview of the ToO system is shown in figure \ref{fig:workflowDiagram}.

\subsection{Triggering alert stream}\label{subsec:alertStream}

Rubin \ac{ToO} observations are initiated upon receipt of an alert from an external experiment through the triggering alert stream, which is distinct from the Rubin alert stream. The Rubin alert stream transmits data containing measurements associated with the detection of a time-variable sources in Rubin difference-images.\cite{DMTN-102, LDM-612} The triggering alert stream, by contrast, ingests and parses data from other physics experiments and forwards them to Rubin infrastructure, where an automated set of \ac{ToO} trigger criteria must be met to execute a \ac{ToO} observation.

All incoming alerts are handled by \ac{SCiMMA} HopSkotch service.\cite{scimma} Alerts from different experiments are assessed for quality after receipt. Alerts that pass the criteria for observation are passed to a database known as the Rubin \ac{EFD}. The application that evaluates alerts and transmits those passing our criteria to the Rubin \ac{EFD} is called the \ac{ToO} Producer. 

\ac{ToO} trigger criteria are defined by the Rubin science community in the latest version of the \ac{ToO} strategy for \ac{LSST}\cite{RubinToO2024}, and are subsequently reviewed by the \ac{SCOC} before adoption as official \ac{LSST} survey strategy. By adopting this approach, future revisions of \ac{ToO} trigger criteria can be developed internal to the ToO producer, separate from the \ac{FBS}, which handles the observing strategy.

\subsection{The Rubin \ac{ToO} Advisory Committee}\label{subsec:ToO_AdvComm} 

The Rubin \ac{ToO} Advisory Board is charged with giving advice on various aspects of the \ac{ToO} process during Rubin Operations. While the triggering of \ac{ToO} observations with Rubin Observatory will largely be by automated algorithms, the \ac{ToO} Board will give input on short ($<24$ hour), medium (days-weeks), and long (months-years) timescales. 

The Board will work directly with the \ac{RTO}, a 24/7/365 role responsible for managing triggers of \ac{ToO}s at the observatory. The \ac{RTO} will be responsible for making  trigger decisions in real time. The Board membership process is run by the Rubin Science Advisory Committee with final approval of invitation to the Board by the Rubin Observatory Director.

The board is made up of experts across the fields of time domain astronomy, including experts in solar system science, \ac{GW} events and their \ac{EM} counterparts, neutrino events and their \ac{EM} counterparts, and \ac{PHAs}. Membership on the board will represent expertise in both scientific and technical aspects of each field. The Board names its own Chair from among its members.

The inaugural \ac{ToO} Advisory committee was formed in November 2025, after the start of operations on 25 October, 2025. Prior to the board's formation, decisions on potential \ac{ToO} triggers were made by members of the Rubin Project commissioning team with expertise in \ac{ToO} science and the observatory's on-sky performance.

\subsection{The Rubin Engineering Facility Database (EFD) and Commandable SAL Components (CSCs)}\label{subsec:EFD}

The Rubin Engineering Facility Database hosts telemetry for all Rubin Observatory components, capturing their performance, status, and telemetry.\cite{DMTN-290} For the purposes of a \ac{ToO} observation, the \ac{EFD} hosts relevant data from \ac{SCiMMA} alert packets. This includes skymap information, alert metadata, and other information required to generate accurate observing strategies.

As part of the \ac{OCS}, every component of Rubin Observatory, such as the telescope mount and \ac{lsstcam}, is controlled by its \ac{CSC}. Observing specialists operate them through ac{SAL} scripts. Each \ac{SAL} can be controlled using a \ac{CSC} by the Rubin observing specialists. For scheduling observations, the \texttt{Scheduler \ac{CSC}} collects telemetry from the \ac{EFD} and passes the information to a driver that formats it in a common schema for the scheduling algorithm. 

By using a general purpose interface to collect and pass telemetry from the \ac{EFD} to the scheduling algorithm, we can easily add new data sources for \ac{ToO} observations. This flexible design supports the \ac{SCiMMA} alert stream, but also alert information added manually by observing specialists or \ac{ToO} scientists for exotic \ac{ToO}s that do not meet standard triggering criteria. \cite{TSTN-035} Manual \ac{ToO} observations are only triggered when an alert does not meet the standard criteria for \ac{ToO} observation, and activation is chosen by the Rubin \ac{ToO} advisory committee.

\subsection{The Rubin Scheduler}\label{subsec:Scheduler}

The \ac{FBS} is a dynamic scheduling system that selects observations by scoring candidate fields based on survey goals and current conditions. It continuously updates priorities using features such as sky brightness, weather, and past coverage to optimize survey performance in real time.\cite{LynnesPaper} The software that encodes the \ac{FBS} is the \texttt{rubin\_scheduler}\cite{fbs} package. The \texttt{rubin\_scheduler} contains a simulation mode to generate simulated observations at high speed, useful for simulating different observing conditions, different survey conditions, and \ac{ToO} alert responses. 

In section \ref{subsec:Scheduler verification}, references to the Rubin Scheduler refer to the simulation mode. In all other sections, mentions of the Rubin Scheduler refer to the \ac{FBS} implementation within the \texttt{Scheduler \ac{CSC}}, the \ac{CSC} responsible for scheduling observations for the \ac{LSST}.

\ac{SCiMMA} alerts passed to the \ac{EFD} have metadata associated with the alert type of the \ac{ToO}. This piece of metadata is passed to the Rubin scheduler to execute the appropriate strategy, as defined in the latest implementation adopted by the \ac{SCOC}.

\subsection{The Prompt Processing Pipeline}\label{subsec:PP}

After the \ac{FBS} generates a \ac{ToO} observing plan and the \texttt{Scheduler \ac{CSC}} executes observations, images are processed using the \ac{LSST} Prompt Processing pipeline to generate alerts. This is not unique to \ac{ToO} observations, as Prompt Processing will run on all science images obtained for the \ac{LSST}. Alerts generated by the \ac{LSST} prompt processing pipeline are \ac{LSST}’s real-time data product derived from image data with no embargo period and data right exclusivity\cite{RTN-114}. 

Within minutes, images are processed and transient sources are measured before data transmission of alerts to \ac{LSST} alert brokers who host the alert data. Within 24 hours, the \ac{PPDB} is updated with the same data as were released in the alerts. Difference-image source linking is run to discover solar system objects. Detections of solar system objects are ingested by the Minor Planets Center and orbital parameters are refined. After 80 hours, the promptly-processed images and difference images become available to users for individual inspection and processing.
\section{Integration to Rubin Observatory}\label{sec:system_integration}
Each aforementioned component of the \ac{ToO} system required integration to the summit environment before a full test and validation of the \ac{ToO} system on sky. Here, we briefly describe the activities to commission the \ac{ToO} system during the on-sky commissioning period in 2025. During this time, the \ac{EFD} had been in use for several years for other data sources, and the prompt-processing pipeline had been in use with other data sources for several years.

\subsection{\ac{SCiMMA} Alert Stream}\label{subsec:alertStream verification}

On June 5, 2025, the \ac{SCiMMA} group demonstrated the capability to send \ac{ToO} alert packets to the Rubin \ac{EFD}, using simulated metadata, skymap information, and alert identifiers. Subsequent work followed to understand specific data definitions, including formatting skymaps as \texttt{NSIDE=32 NESTED} ordered \texttt{healpix} maps.\footnote{\href{http://healpix.sourceforge.net}{healpix.sourceforge.net}}

Additional work focused on integrating gravitational wave alerts from \ac{LVK} to the \ac{ToO} producer as the first alert type. The first alert received at the Rubin \ac{EFD} was a simulated \ac{GW} alert on June 5th, 2025. Additional alert types have been integrated to the \ac{ToO} producer, and high-energy neutrino and galactic supernovae alerts are now supported for automated alert processing and \ac{ToO} observation. Development in support of \ac{PHAs} is ongoing. 

\subsection{The Rubin Scheduler}\label{subsec:Scheduler verification}

While the \texttt{rubin\_scheduler} has been in use at Rubin Observatory with the Auxiliary Telescope\cite{auxtel_commissioning} since 2022, and with \ac{lsstcomcam} in late 2024 \cite{LSSTComCam}, the \ac{ToO} usage was never used with either telescope or instrument. 

Each alert type has different observing strategies defined by the Rubin science community and endorsed by the \ac{SCOC}. To evaluate the efficacy of these observing strategies, we performed tests on simulated \ac{ToO}'s using the \texttt{rubin\_scheduler} software package under relevant alert, sky, and \ac{lsstcam} filter conditions.

To test each \ac{ToO} strategy, an \ac{LSST} survey configuration was injected with a \ac{ToO} of the relevant type and skymap characteristics. The simulation of the \ac{LSST} survey was then enabled to run, such that the \ac{ToO} was detected by the scheduler, which planned and executed the \ac{ToO} observations following the encoded observing strategy.




All \ac{ToO} strategies passed validation tests in the \texttt{rubin\_scheduler} and were deemed adequate for on-sky use. Several caveats to observing strategies were identified related to environmental conditions for binary black-hole GW follow up. At this time, we did not evaluate the sensitivity of the \ac{ToO} observing strategies, and an analysis of the existing observing strategies is ongoing\cite{RTN-107}.


\subsection{Full system tests}\label{subsec:systemTests verification}


To test the \ac{ToO} system from end-to-end on-sky, we injected a simulated \ac{ToO} event in a region of the sky where template images existed. We triggered this \ac{ToO} on the night of July 6, 2025, to evaluate the system readiness, response latency, and image processing outcomes. We focused on an area centered at (285.2,-19.8), known as the New Horizons field. 

The injected alert schema was identical to the \ac{LVK} AVRO schema, shifted to center on the New Horizons field. To simulate an object on the new horizons field, we ran simulations of \ac{BBH} \ac{GW} events in the region, and modified several parameters in the alert packet to pass the alert criteria thresholds, including the \texttt{HasNS}, \texttt{HasRemnant}, and \texttt{classification} fields. 

The observing strategy was configured using 2 visits per epoch, epochs spaced at  0, 1, 2, and 24 hours, griz bands for the first three epochs, ri for the final epoch, 30 second exposures, and masks to preferentially avoid the moon and regions of high wind. The bands were chosen to evaluate the impact of filter change cadence on the resulting \ac{ToO} observations.

The alert was manually injected to the \ac{SCiMMA} alert stream at 03:22:54 UTC on July 7 2025, and received at EFD two seconds later. The alert was processed by the Scheduler \ac{CSC} at 03:32:07 UTC, shortly after the \ac{ToO} scheduler configuration was enabled, first observations of the field began at 03:33:47.


This test was deemed a success, as the \ac{ToO} alert interrupted planned observations without human intervention. Observations were scheduled appropriately according to the injected \ac{ToO} observing strategy, and the requested bands were used. Due to errors with the \ac{FES} during filter changes, z band observations were not acquired for the first epoch, though they were scheduled before errors ocurred. The \ac{ToO} response did not interrupt the next set of exposures, but this choice was deliberate to evaluate staged \ac{ToO} exposures before execution. A configuration change was made to the Scheduler \ac{CSC} to interrupt the scheduled observations for all future \ac{ToO} observations.

For this test, we elected to observe only the first two epochs due to constraints from other commissioning activities. Observations were concluded at 05:27:59 UTC. 

At the time of this test, template images used for difference imaging we not acquired by Rubin in a wide area of the sky. For this reason, the New Horizons field was the optimal choice for comparing the performance of a custom image processing pipeline using \ac{DECam} templates, as Rubin had already acquired templates in multiple bands in this region. The commissioning team prepared a difference imaging pipeline using the LSST Science Pipelines\cite{PSTN-019} that makes use of previously coadded images acquired using \ac{DECam} from the DES, DELVE, and DECADE surveys. \cite{DES_dr2,DELVE_dr2,DECADE_release}

Due to image processing constraints and focus on other commissioning activities, image processing was ultimately never pursued for this field.






\section{Performance during Rubin Observatory early operations}\label{sec:commissioning_performance}
\subsection{Interstellar Comet 3I/ATLAS}\label{subsec:3I}

Interstellar Comet 3I/ATLAS was discovered on 1 July 2025 by the ATLAS survey telescope at Río Hurtado, Chile \cite{seligman2025discoverypreliminarycharacterizationinterstellar}. It is the third interstellar object confirmed passing through the Solar System, after 1I/'Oumuamua (discovered 19 October 2017) and 2I/Borisov (discovered 29 August 2019). Given the rarity of the object and the potential for high-impact science, Rubin-commissioning \ac{ToO} advisory board elected to test the \ac{ToO} system with 3I/ATLAS. 

Many other science exposures of 3I/ATLAS had been obtained before \ac{ToO} activation, as the object was in the Rubin SV survey footprint for $\sim2$ weeks before its identification. The lack of previous templates in the region prevented earlier observation and analysis, despite detection in Rubin images as early as 2025-06-24.\cite{3I_rubin_paper} Here, we focus only on observations triggered by the \ac{ToO} system.

After consultation with the Rubin solar-system science unit, we devised a custom observing strategy for 3I/ATLAS that would maximize the probability of extracting a complete color profile of 3I/ATLAS, and determine its rotational period. The observing plan was defined to use a single \ac{ToO} epoch with 30 second exposures, repeat exposures four times per pointing, apply a maximum dither of 0.01 $\deg$ between each exposure to capture any area that was in chip gaps of the \ac{lsstcam} focal plane, and use an identical suite of safety masks at the SV surveys. Since the localization area of 3I/ATLAS was extremely small ($<1\deg^2$), the injected alert localization was refined to fit within the nearest single $\texttt{NSIDE=32}$ flattened healpix pixel, the format utilized by the Rubin scheduler.
\subsubsection{Observation Latency}
The alert was injected to the SCiMMA alert stream at 2025-07-13 2:39:12 UTC, and promptly received at the Rubin EFD within one second of injection. The deployed configuration of the \texttt{Scheduler \ac{CSC}} did not immediately clear the observation queue, mainly to ensure that the requested \ac{ToO} observations were correct ahead of on-sky execution. This delayed the start of \ac{ToO} observations of the 3I/ATLAS field began until 3:09:13 UTC, starting with observations in g band. Observations were completed by 04:44:30 UTC, and we obtained detections of Interstellar Object 3I/ATLAS in all five available \ac{lsstcam} bands (grizY). 

\subsubsection{Areas of Improvement}
Despite detection of our intended transient, we identified two issues during \ac{ToO} execution:

\paragraph{Filter change behavior:}\label{subsubsec:changeBehav}

With four visits in the scheduler configuration, the expectation for observations was to complete four visits per pointing in each filter, before changing the filter and proceeding with observations. In on-sky testing, we obtained a single observation in all five filters, before repeating the tilings an additional three times. This repetitive cycle of four identical tilings appeared reflective of the configured visits, but not optimized to complete observations in each band before changing filter. 

The time to change filter is $\sim90$ seconds, so this cadence resulted in an additional 15 filter changes, or $\sim22.5$ minutes of filter change time that could not be utilized on sky. To address this issue, we changed the scheduling algorithm to prioritize observations in a given band before changing filter.

\paragraph{Visit locations:}

The localization region supplied to the Rubin scheduler for this ToO alert was one single healpixel, closest to the known position of 3I/ATLAS. Under the deployed observation tiling method, the boresight-pointing grid remained fixed from its start-of-night generation, and can have multiple pointings inside a single $\texttt{NSIDE}=32$ pixel. Based on the repetitive behavior of exposures, we determined that the scheduler identified two pointings inside the healpix pixel, and imaged them twice, once with the specified dither, to reach four visits in each band.

This pointing generation is not optimal, as a single \texttt{NSIDE=32 HEALPIX} pixel has an area of $\sim3.3 \deg^2$, or $\sim1/3$ the FOV of \ac{lsstcam}. As an intermediate solution, we deployed changes to define a pointing grid for \ac{ToO}s that centers pointings on each \texttt{HEALPIX} pixel. This change allows us to better ensure that all visits for well-localized \ac{ToO}s are adequately covered, without risking vignetting of desired regions of the sky.

\begin{figure}
    \centering
    \includegraphics[width=.8\linewidth]{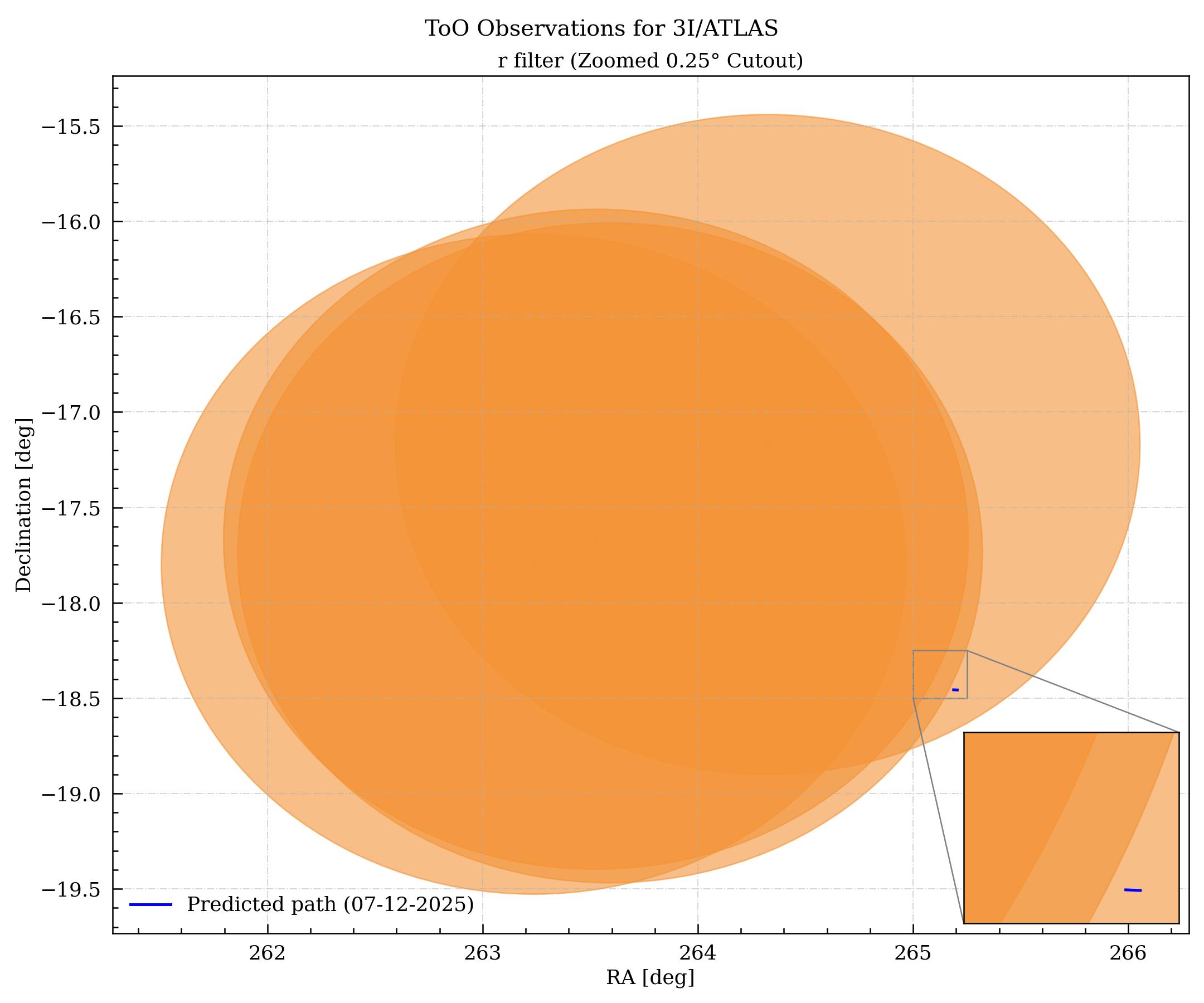}
    \caption{The Rubin pointings for the 3I/ATLAS \ac{ToO}, with four distinct pointings in the region of 3I/ATLAS, but only one pointing overlapping with the resulting position. Predicted position is using the Horizons On-Line Ephemeris System and previous detections of 3I/ATLAS from the Asteroid Terrestrial-impact Last Alert System\cite{2025MNRAS.542L.139B}.
    }
    \label{fig:3I_atlas}
\end{figure}

\subsection{GW Candidate S250725j}\label{subsec:S250725j}

While Rubin Observatory was commissioning, \ac{LVK} was in the midst of its longest observing run, \ac{O4}, which lasted from 2023/05/24 - 2025/11/18. During \ac{O4}, \ac{LVK} was regularly sending alerts about \ac{GW} candidates through Kafka streams, including \ac{SCiMMA}. On 2025-07-25 at 04:09:58 UTC, the gravitational wave candidate S250725j was observed by the \ac{LVK} detector network. S250725j is a high-significance \ac{BBH} event, localized to a 90\% area of 18.6 $\deg^2$. The area localization of the candidate met the alert quality criteria for a \ac{BBH} event. After discussion within the Rubin-commissioning \ac{ToO} advisory board, the decision was made to trigger a \ac{ToO} on S250725j.

The observing plan was devised to follow the \ac{BBH} observing strategy as laid out in the latest endorsed observing strategy from the \ac{SCOC}. This was designed to use 30 second exposures, one visit per epoch, with epoch spacing on nights 1-3-8-10-39 after the alert is received, a dither between exposures with a maximum of 0.01 degrees (36 arcseconds), and the complete suite of safety masks used by the SV surveys. The requested filters for a \ac{BBH} merger of this distance and type is u-g-r bands. Due to an engineering constraint with the \ac{FES} of \ac{lsstcam}, the g band was not available. As a result, we set the observing strategy to observe in u-r-i bands. Observations began on the night of 2025-07-28, and were the only observations obtained of the region of interest, due to a combination of severe weather constraints and other commissioning priorities.
\subsubsection{Observation Latency}
The alert packet was submitted to the ToO alert stream on 2025-07-29 at 02:16:22 UTC, and received at the Rubin EFD at 02:16:25 UTC, approximately 3 seconds later. Observations began at 02:28:07 UTC, and concluded at 03:22:58.
\subsubsection{Observation Depth and Coverage Efficiency}
\begin{figure}[t]
    \centering
    \includegraphics[width=0.49\linewidth]{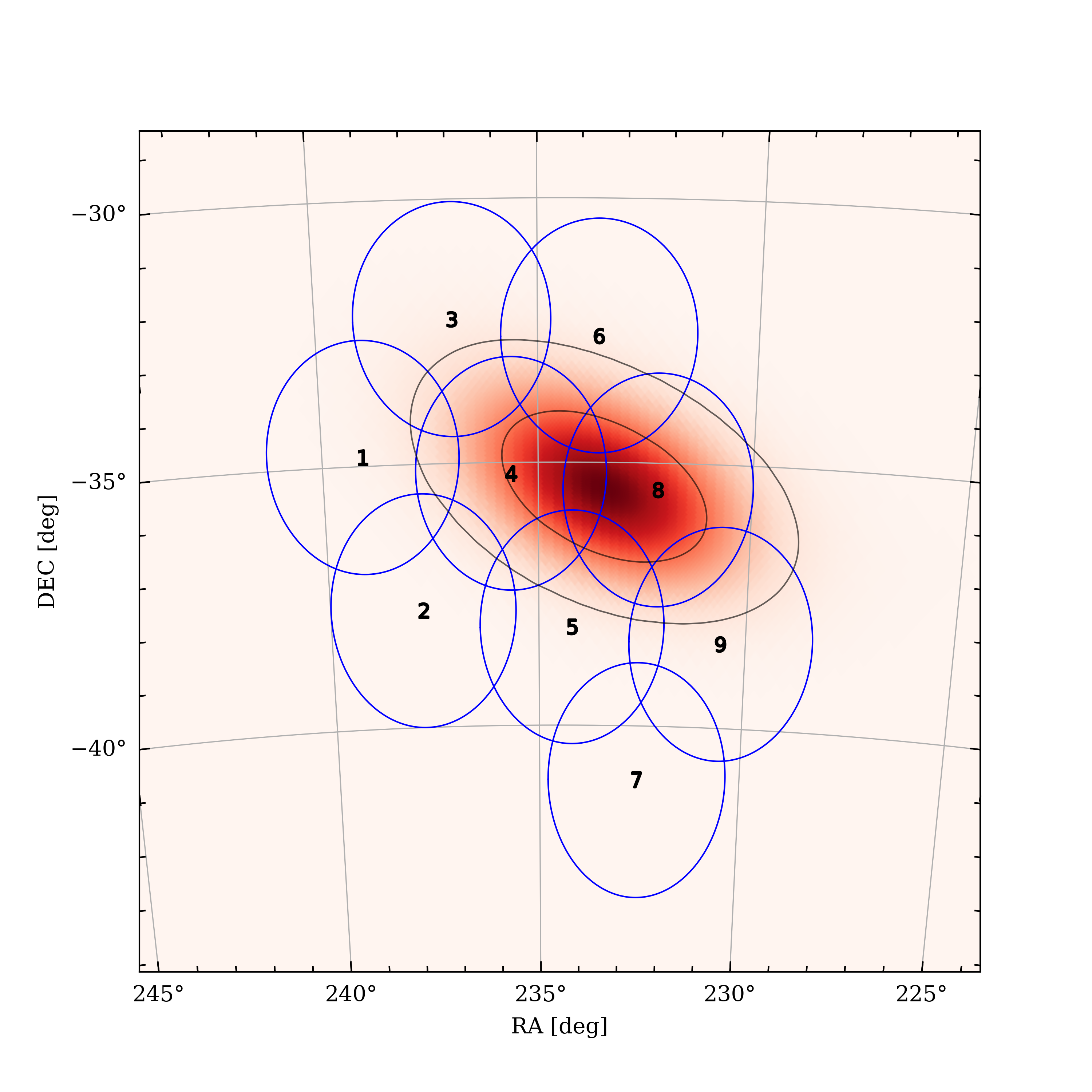}
    \includegraphics[width=0.49\linewidth]{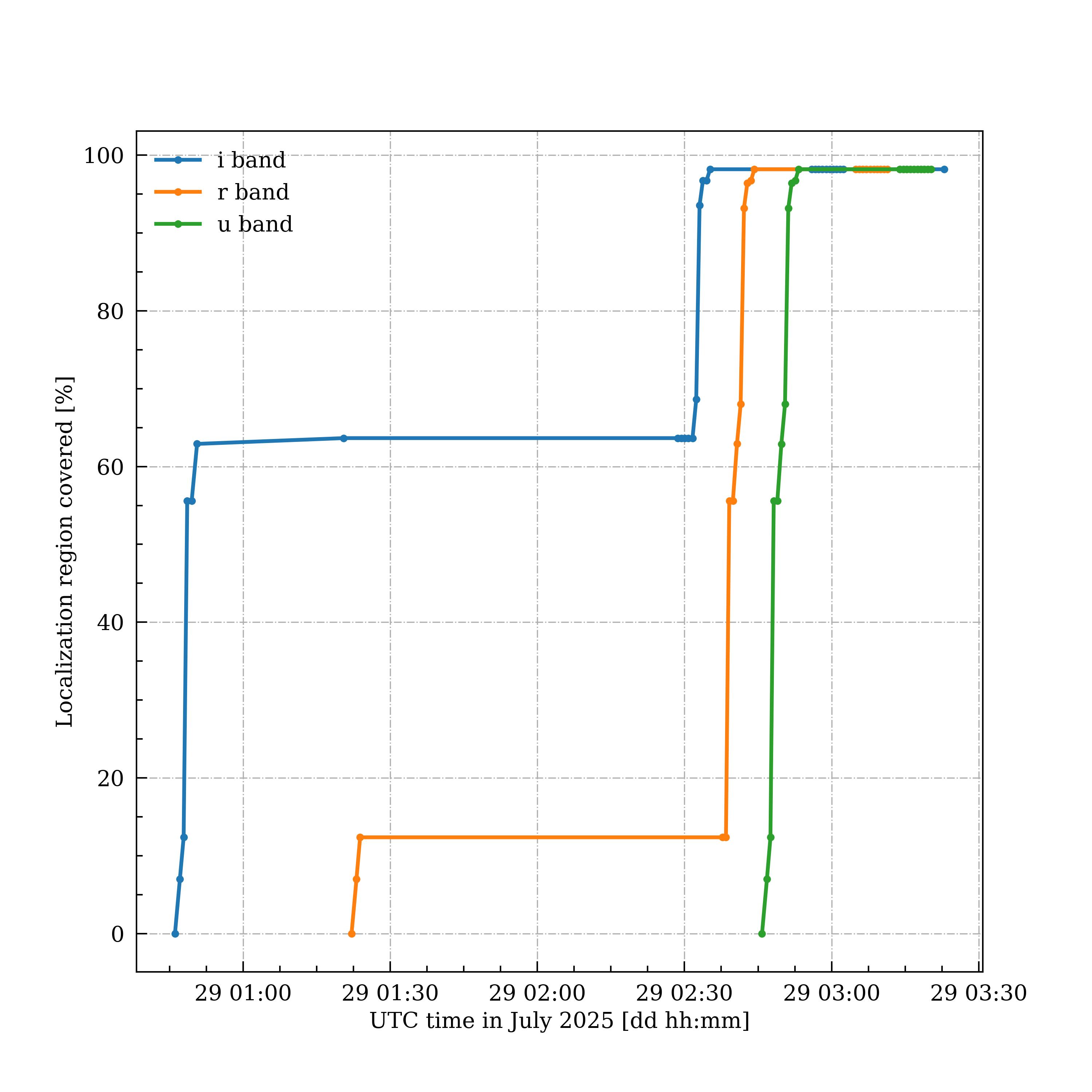}
    \caption{Left: The localization skymap associated with S250725j, with the executed Rubin pointings overlaid. The 50\% and 90\% contours are shown on the localization probability, along with the numbering of each pointing. 
    Right: The cumulative probability coverage of S250275j on the night of July 28th, 2025, separated by band.}
    \label{fig:S250725j}
\end{figure}

Conditions on the night of observing were challenged by atmospheric seeing and proximity of the region-of-interest to the galactic plane, which led to variable image quality of 1-1.8" PSF. We reached a median depth of 24.1 mag in r band, 23.8 mag in the i band, and 23.6 in u band, which was the deepest wide-area ($\Omega_{90}>10\deg^2$) search for an \ac{EM} counterpart to a \ac{GW} candidate to date. An issue related to the Scheduler CSC caused the field to be observed two times, which was similar to the filter-dither prioritization behavior noted in section \ref{subsubsec:changeBehav}, but tracked to a different issue independent of the \ac{ToO} system. 
The changes to the pointing generation and filter-visit priority were deployed to summit, and manifest through the ordered spacing of pointings, shown in figure \ref{fig:S250725j}. 

This ToO is the first test on-sky of a ToO area that is larger than a single Rubin pointing. Therefore, this event provides a unique opportunity to evaluate the efficiency of covering a ToO area to completion.
For each pass in each filter, 100\% of the 50\% probability region was covered, and 99.576\% of the 90\% region was covered. Additionally, the 50\% region was covered at a rate of $1.64 \deg^2$/min, and the 90\% region was covered at a rate of 4 $\deg^2/$min.

\subsubsection{Image Processing and Analysis}

To process and analyze the observations from this event, we used an image processing pipeline as described in section \ref{subsec:systemTests verification}, using previously coadded images acquired using \ac{DECam} from the DES survey. The location of this \ac{GW} candidate is nearby the galactic plane, which led to known problems with the WCS solution. The issues with the WCS solution in search images propagated to difference images, resulting in many bad subtractions, dipoles, and other image artifacts.

\begin{figure}[b]
    \centering

    \begin{minipage}[c]{0.39\textwidth}
        \centering
        \includegraphics[width=\linewidth]{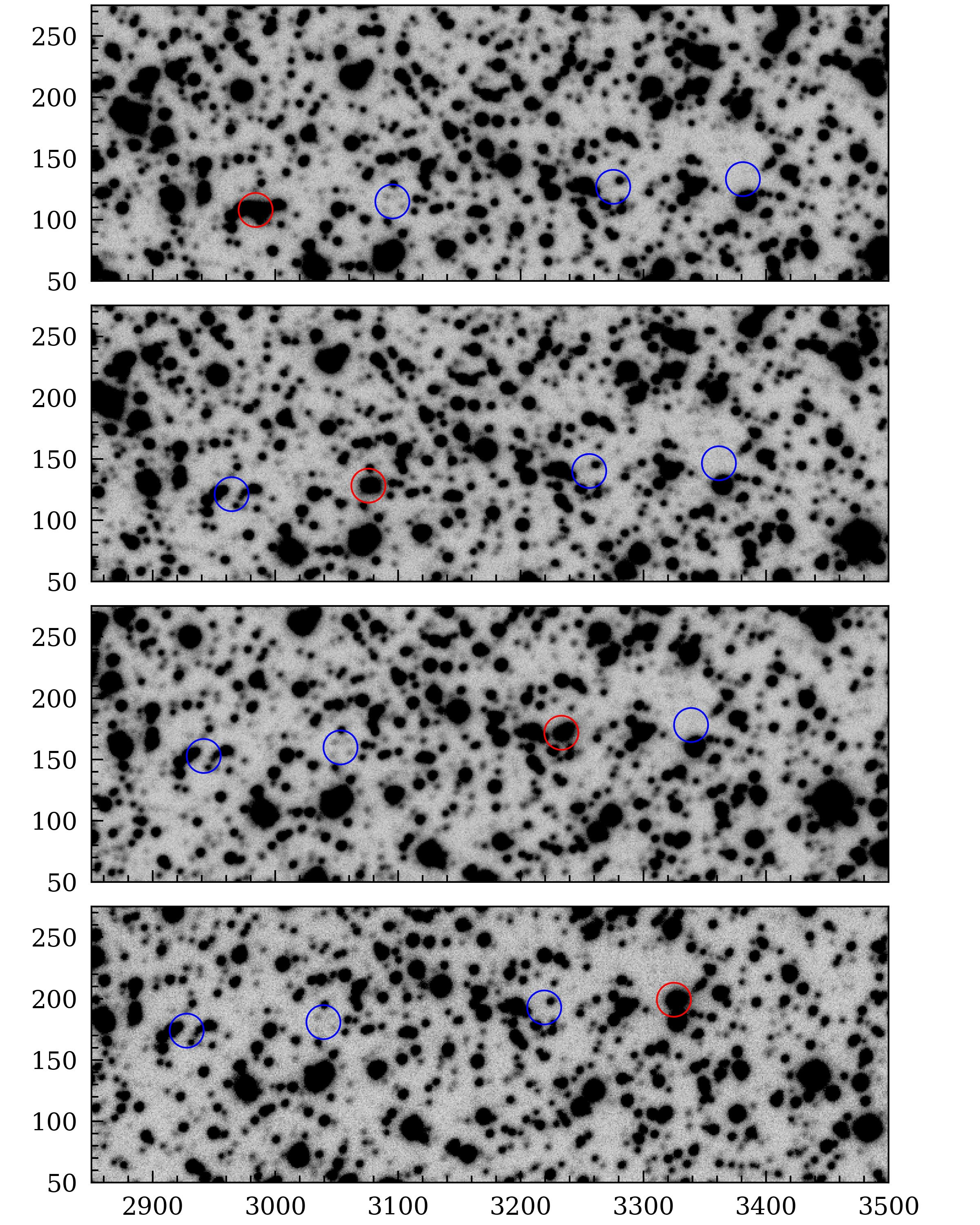}
    \end{minipage}
    \hfill
    \begin{minipage}[c]{0.59\textwidth}
        \centering
        \includegraphics[width=\linewidth]{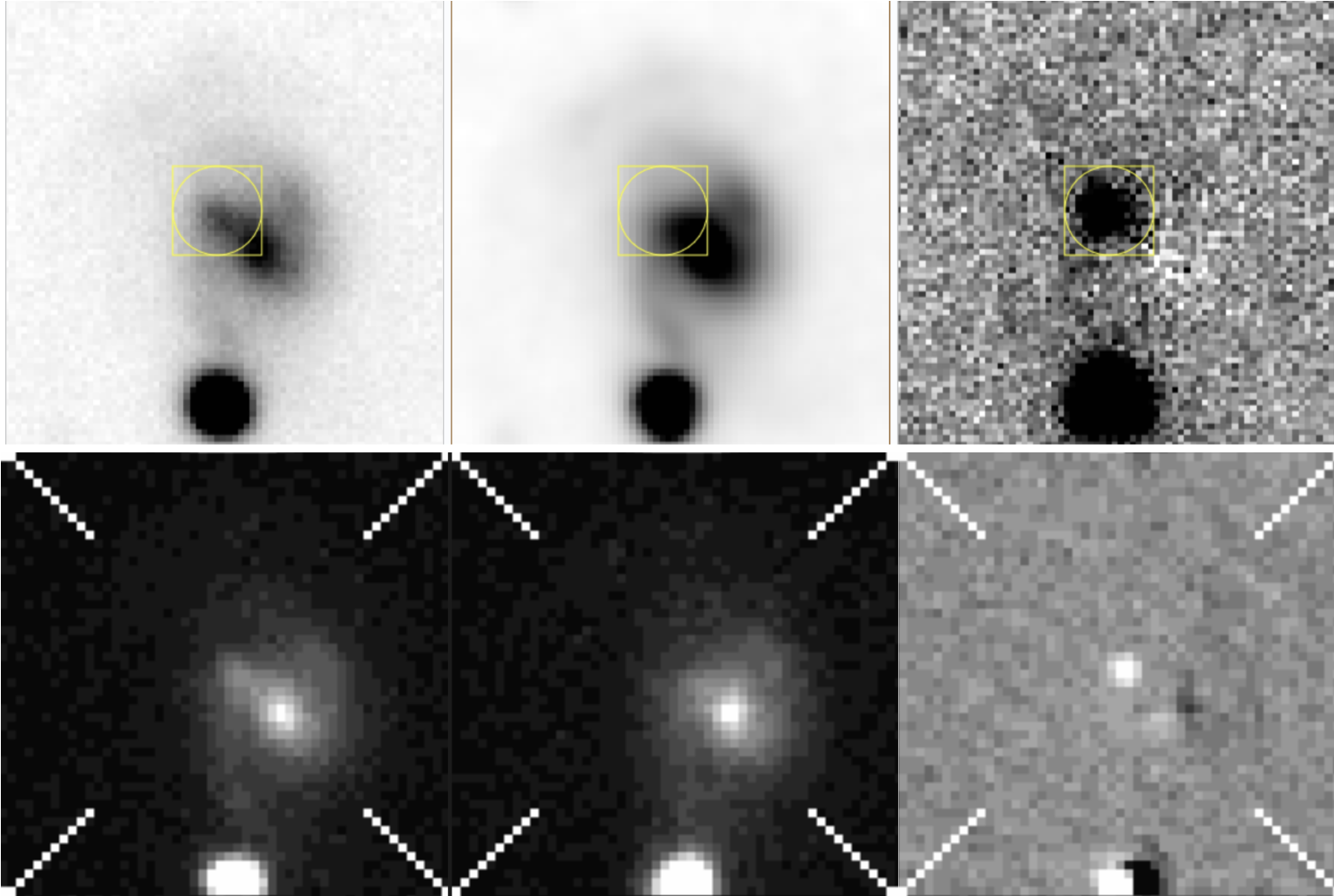}
    \end{minipage}
    \caption{Left: Images of interstellar comet 3I/ATLAS from a cutout of a single \ac{lsstcam} CCD. The position of 3I/ATLAS for each exposure is shown in red, with the positions in the other exposures marked in blue. Coordinates are in detector frame. Right: AT2025sib, shown in the yellow reticle. One of the four reported Rubin candidates from the S251112cm followup,\cite{S251112cm_0} as viewed by LSSTCam (top) and DECam (bottom), each with a field of view of ~14” on each edge. Search (left), template (center), and difference (right) images.}
    \label{fig:candidates}
\end{figure}

After processing the relevant visits, we applied a minimal set of flags to remove visits that have dipoles, cosmic rays, and other artifacts. Despite this quality cut, we identified $\sim25$k DIA candidates per visit, which far exceeds the ability of human vetting capabilities. Due to a lack of prior and subsequent Rubin observations, constraint on temporal evolution of the sources was not available. For this reason, the commissioning analysis group pivoted to review of transient candidates reported by other teams using DECam\cite{S250725j_0, S250725j_1} (see figure \ref{fig:candidates}). The results of the inspection of \ac{DECam} candidates were reported in a \ac{GCN} circular.\cite{MacBride_Anand_Howard_2025}

\subsection{\ac{GW} Candidate S251112cm}\label{subsec:S251112cm}

In November 2025, the \ac{ToO} advisory board was onboarding to their role in Rubin Observatory operations. This timing proved opportune, as on 2025-11-12 at 15:19:32 UTC, the gravitational wave candidate S251112cm was observed by the \ac{LVK} detector network. S251112cm was designated as a high-significance, \ac{SSM} candidate, the first publicly reported by \ac{LVK}. S251112cm was initially localized to a 90\% area of 1,220 $\deg^2$. Despite the large localization area, the \ac{ToO} advisory board endorsed a \ac{ToO} trigger on S251112cm, due to the opportunity for discovering an \ac{EM} counterpart from a nearby source (posterior mean distance of 92 Mpc inferred from the \ac{GW} signal). This \ac{ToO} represented the first on-sky demonstration of Rubin Observatory focusing on wide-area discovery of a faint optical counterpart.

In November 2025, Rubin Observatory was in an early operations period of continued system optimization, prior to beginning sustained \ac{LSST} observations. The observing strategy for this event was designed to image the accessible portion of the 70\% localization area in g and i bands, with one 30 second visit for each band, on nights 0, 2, 4, and 6 of the observing campaign. The decision to focus on the 70\% localization area instead of the standard 90\% localization area was to balance discovery opportunities with ongoing commissioning activities.

\subsubsection{Observation Latency}

Due to the early-operations environment of November 2025, the \ac{ToO} producer, which is responsible for assessing the quality of \ac{ToO} alerts for Rubin observation, was not running at the time of the \ac{LVK} alert publication. Additionally, the trigger criteria would not be met based on the contents of the alert packet and the in-place alert criteria. This combination of factors meant that the \ac{ToO} alert packet needed to be manually submitted to the external trigger stream, which was done at 2025-11-15 00:53:52 UTC. The alert was received at the Rubin \ac{EFD} one second later. First observations of the localization area of S251112cm began at 01:49:27 UTC, after delays due to deployment of a scheduling configuration that utilized non-standard exposure times. The remainder of S251112cm observations utilized the standard 30 second exposure times.

\subsubsection{Observation Depth and Coverage Efficiency}

On November 15 2025, at 01:49:27 UTC, we began observing the localization region of S251112cm, using the initial localization information provided by \ac{LVK}. On the first night of observation, we observed 39.2\% ($849.1 \deg^2$) of the localization area in the g-band, and 33.5\% ($754.9 \deg^2$) of the localization area in the i band. The difference in coverage was due to the localization area of S251112cm setting early in the night in Chile, before the telescope could observe the region in i band. 

This lack of initial area in i-band was subsequently observed the following night, demonstrating that the \ac{ToO} object was retained according to its specified retention time (48 hours for each observation epoch). Subsequent observations on the nights of November 16-22 mimicked this behavior, observing a majority of the $\sim850\deg^2$ region in Rubin g-i bands on the first night of each night pair (ex., November 16, treated as night 2 after \ac{ToO} alert submission), and compensating for unobserved areas in the following night (ex., November 17, treated as night 3 after \ac{ToO} alert submission).

This \ac{ToO} provided the first opportunity to evaluate the efficiency of Rubin Observatory on sky, in wide-area discovery mode. Rubin Observatory imaged $\sim40\%$ of the S251112cm localization in $\sim3$ hours in two bands, covering $\sim850\deg^2$ in each band and obtaining $\sim100$ images per band for each observation epoch. The coverage speed of the \ac{ToO} field of observation was 10.3 $\deg^2 \text{min}^{-1}$ across all bands. Over the entire \ac{ToO} observing campaign, we achieved a median $5\sigma$ depth of $m_g\sim24.46$, $m_i\sim23.53$, comprising the deepest wide area search for an electromagnetic counterpart of a \ac{GW} event to date.

\begin{figure}[t]
    \centering
    \includegraphics[width=0.49\linewidth]{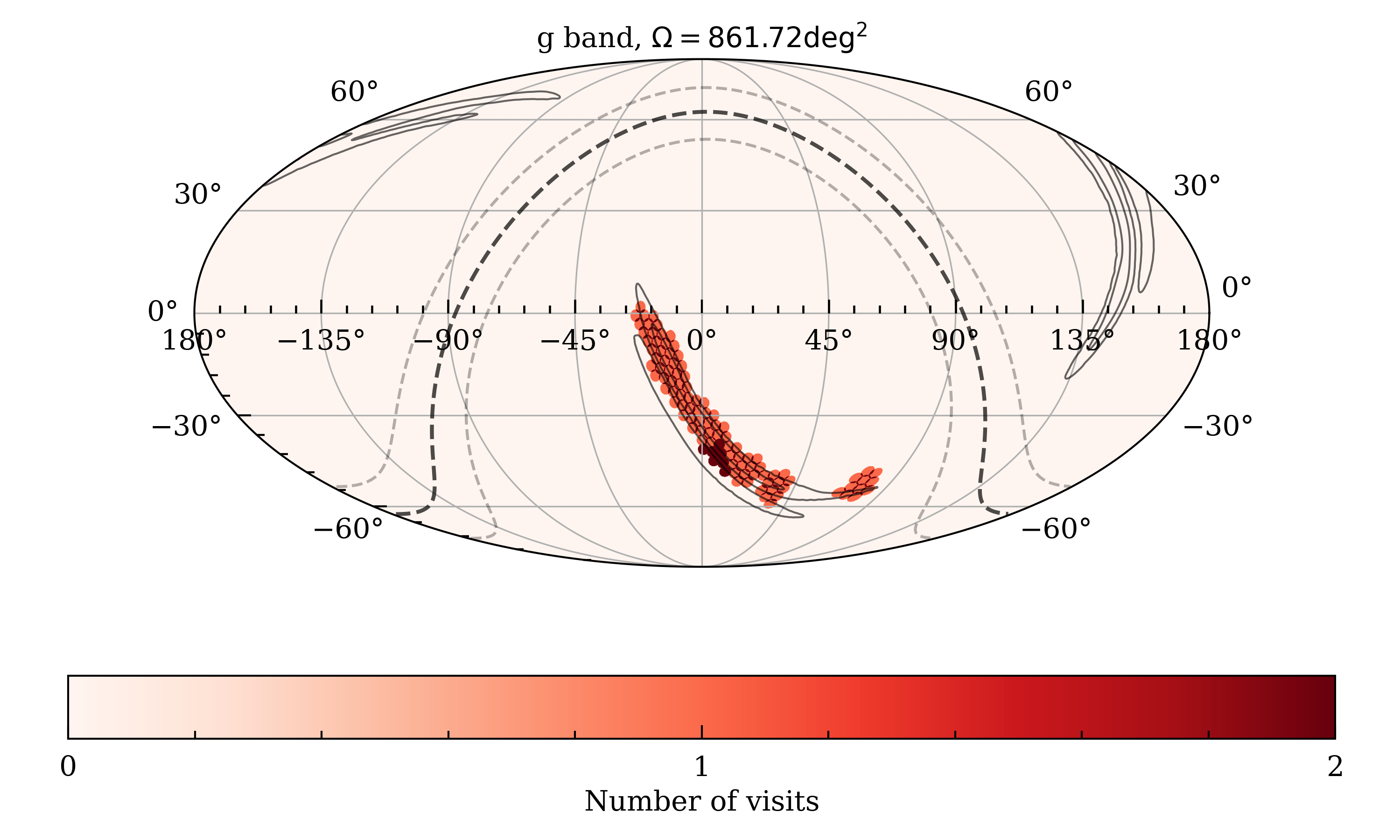}
    \includegraphics[width=0.49\linewidth]{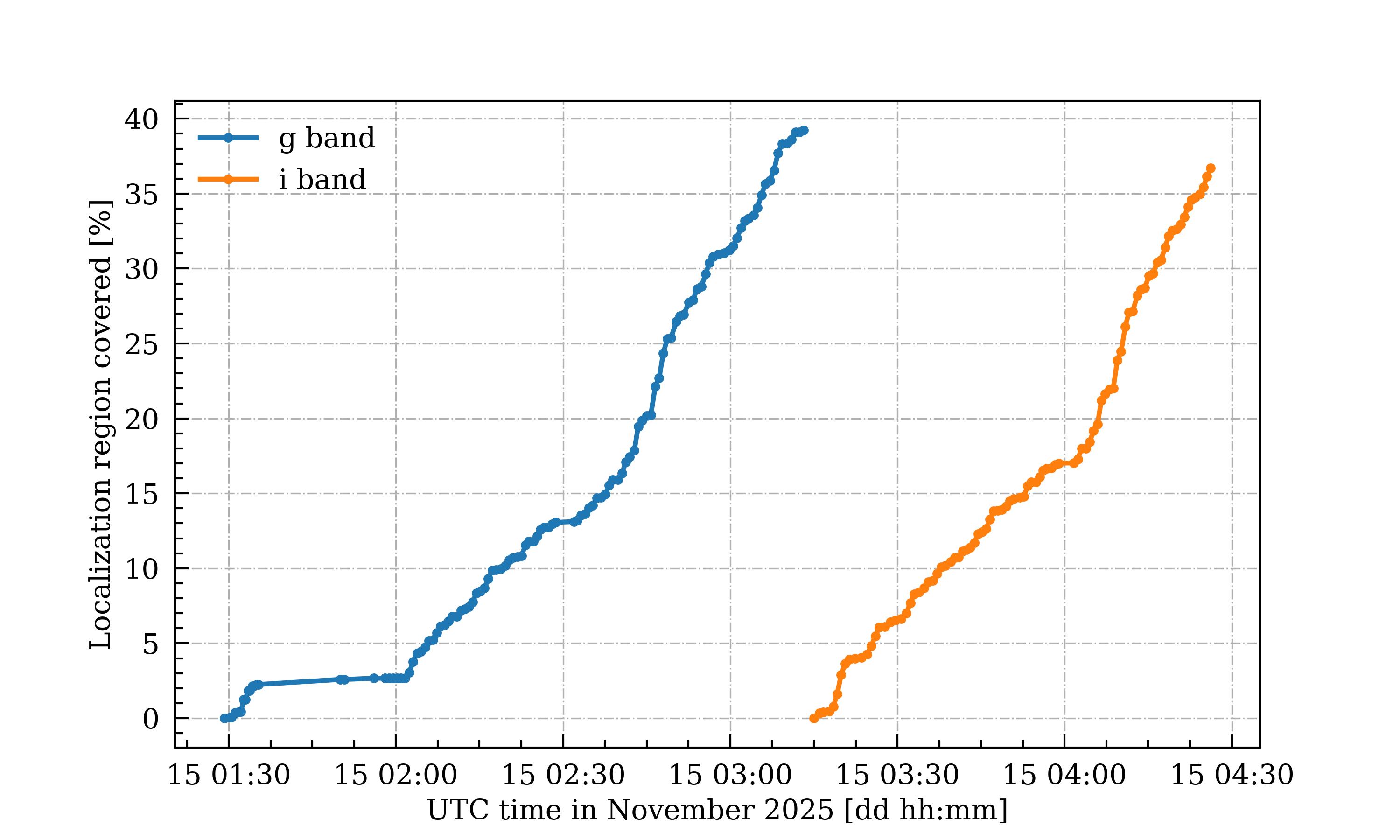}
    \caption{Left: The visits in i band on the night of November 14, 2025. The total area covered on the night in g band, related to the \ac{ToO} response, is 861$\deg^2$, with some fields revisited. The 50\% and 90\% contours are shown in solid line, and the galactic plane and a 15\textdegree offset are shown in dashed lines. Right: The cumulative localization probability coverage of Rubin Observatory in response to the \ac{SSM} \ac{GW} merger candidate S251112cm, separated by g and i band, constrained to the night of November 14, 2025.}
    \label{fig:S251112cm_obs}
\end{figure}

\subsubsection{Image Processing and Analysis}

A small region of the covered area overlapped with the ELAISS1 \ac{DDF}, and generated alerts were sent internally to \ac{LSST} alert brokers, which were also in a commissioning period. For the rest of the imaging area, we performed image differencing against \ac{DECam} templates. Apart from the \ac{DECam} template task, we utilized the standard \ac{LSST} data release production pipeline to obtain measurements by subtracting later Rubin epochs from templates constructed from the first Rubin observation. An on-call team of Rubin Project members with expertise in transient science performed visual inspection on difference images coincident with previously reported candidates from other teams. Many candidates were identified during visual inspection, and reported in two \ac{GCN} circulars.\cite{S251112cm_0,S251112cm_1} During analysis, we observed spatially correlated artifacts related to the DECam template creation, which contaminated many difference object detections. 

In addition to the circulars, forced photometry for high-confidence candidates was reported to \ac{TNS} for community followup and analysis. Further analysis of these observations is ongoing\cite{S251112cm_analysis}.

\section{Lessons learned}\label{sec:lessons_learned}
The first activations and testing of the Rubin ToO program have been successful in demonstrating both technical readiness and scientific capabilities. With the achievements and difficulties in mind, there are several lessons to apply to the \ac{ToO} program during the ten year \ac{LSST}.

\begin{enumerate}




\item \textbf{Consolidation and Availability of Alert Sources}

Rapid and reliable ToO response requires alerts to be ingested in machine-readable form. Machine readable alerts significantly reduce latency compared to manual workflows. Of the four experiments used for Rubin ToO triggers, \ac{LVK} (\ac{GW}) and IceCube (\ac{HEv}) alerts are fully supported through a Kafka stream on SCiMMA. For galactic supernovae, \ac{Super-K} alerts are supported as a mirror of a GCN kafka stream with no integration to \ac{SNEWS}. Additionally, the \ac{SNEWS} alert packets do not contain any localization information, and work is ongoing to directly integrate \ac{Super-K} alerts to Rubin infrastructure through \ac{SCiMMA}.

As of \today, no kafka stream for potentially hazardous asteroids is available. We encourage the \ac{JPL}-Scout program and the forthcoming \ac{JPL}-NEO Surveyor program to publish fully-machine-readable alerts to a Kafka stream, to support localization efforts from Rubin \ac{ToO} follow-up. 

\item \textbf{General Purpose Alert Schemas for the Rubin \ac{EFD}}

The Rubin \ac{ToO} program spans a wide range of astrophysical phenomena, with distinct metadata for each alert. A general purpose alert schema at the Rubin \ac{EFD} allows the Observatory to support existing science cases with the required metadata, and new science cases without requiring disruptive changes to the core scheduling or control software.

The current schema is supportive of probability-weighted distance estimation used in GW followup, and extinction priors for galactic supernovae observations. This level of detail, remaining concise yet informative, provides the necessary information for observing strategies to be exact in their execution.

\item \textbf{System Readiness Metrics}

\ac{ToO} readiness and analysis of a response must be assessed using metrics that extend beyond image-quality conditions. System availability, alert ingestion latency, scheduling response time, and interaction with telescope operators are equally critical for successful \ac{ToO} execution.

Establishing standard metrics for every ToO response provides an active assessment of the Observatory’s ability to respond to time-critical science and identifies bottlenecks that may not be visible in traditional observing efficiency statistics. For future \ac{ToO}s, we envision automated logging of a specific set of metrics relevant to \ac{ToO} science to better inform the science community about the performance and limitations of the system.

\item \textbf{Clear Definition of Operational Workflow and Coordination}

One source of friction during SV \ac{ToO}s was that many individuals had the expertise to contribute to a \ac{ToO} campaign, but no single person or entity had the authority to adjudicate and commit to a specific direction. While operations will benefit from a \ac{ToO} advisory committee, codifying the decision-making process for non-automated responses remains a priority.

As we transition to operations, a clear technical workflow similar to figure \ref{fig:workflowDiagram} needs augmentation with decision nodes for relevant stakeholders, whether human or machine. In addition to workflow definition, the appropriate authority must be granted to stakeholders at their respective decision nodes. This will establish clear responsibilities across summit operations, data management, and science teams. Formal coordination will minimize ambiguity during high-pressure events and ensure that operational decisions are consistent, auditable, and reproducible. 

\item \textbf{Communication Channels with Summit Operations Staff}

Providing clear communication to summit operations staff is crucial for the long term success of the Rubin \ac{ToO} program. Drawing on the experience from the observing campaigns during the SV program, two areas of improvement are necessary:

\begin{itemize}
    \item \textbf{Communication of alert receipt to summit staff:} Currently, a slack bot is the only communication mechanism visible on public communication channels. An alarm in the \ac{OCS}, using the \ac{LOVE} system\cite{LOVEPaper} would be the most direct way to immediately notify the relevant summit staff that a \ac{ToO} has entered the observation queue, and development is ongoing to implement an alert mechanism to on-call summit staff.
    \item \textbf{Simulated observing plans and details for received ToO alerts:} This resource will provide observers with the foreknowledge to anticipate changes to the observing schedule, including expected observation times of specific \ac{ToO}s. For automated alerts, development is ongoing to provide simulated observation projections to summit operations staff.
\end{itemize}

This transparency reduces operational stress and improves overall ef1ficiency during time-critical campaigns.

\item \textbf{Defined \ac{ToO} Campaign Outcomes}

Clearly articulated success criteria for \ac{ToO} targets serve three purposes:

\begin{itemize}
    \item Quantitative assessment of a specific observing campaign - did we observe too much, or too little? With these answers in hand, future revisions to observing strategy can be grounded in data and historical outcomes. 
    \item By defining the conditions where a target has been sufficiently observed, processed, and analyzed by Rubin and handed off to other observatories, we can inform the decision to stop observing a \ac{ToO} and return to observing other fields of the \ac{LSST}.
    \item These critera enable transparent reporting to the broader community about the success or otherwise of a specific \ac{ToO} campaign.
\end{itemize}

Potential success criteria could be as simple as detection in a single band, or as complex as coverage of 24 magnitude or greater in r band over the entire 90\% localization region within eighteen hours of alert receipt. These success definitions, specific to each \ac{ToO} target class and informed by experience from both operations and the science community, will guide \ac{LSST} through data-driven revision of observing strategies and criteria-driven decisions on observing plans. 

\item \textbf{Interventional Measures to Start and Stop Observations}

We cannot predict when exotic phenomena will occur over the ten-year \ac{LSST}. As demonstrated in commissioning, a non-standard \ac{ToO} (3I-ATLAS) occurred, requiring a \ac{ToO} response outside of the standard operation. The ability to initiate \ac{ToO} observations outside of the standard alert criteria, while bespoke, exists in the current implementation. Similarly, operational controls must support the intervention to terminate \ac{ToO} observations once success has been achieved. This mechanism provides a safeguard against erroneous alerts and enables a flexible response to evolving science priorities.

In addition to developing the technical capability, the workflow from the authority-holding person or entity to the personnel with the technical expertise to execute on specific decisions must be defined. Whenever possible, lower barriers to system interaction will aid in minimizing the latency of decisions propagating to the system.

\item \textbf{Data Infrastructure Stability During Image Processing}

Rubin \ac{ToO} science depends on the stability and uptime of the \ac{usdf}. While responding to the S250725j trigger, some services including \texttt{Firefly} were unreliable, preventing rapid analysis of potential candidates. Additionally, bespoke image processing was delayed due to standard queue wait times for Rubin users on \ac{usdf}. 

Delays in image processing due to long job queue wait times were significant, and hindered the timely dissemination of high significance Rubin candidates to the community. If possible, a special accounting group for time sensitive \ac{ToO} reprocessing of Rubin image products would aid significantly in rapid analysis and information dissemination of specific candidates to the scientific community.


\item \textbf{Clear Communication and Engagement with Other Experiments}

\ac{ToO} operations are multi-messenger - both from the science perspective and the operational perspective. Formal communications with \ac{LVK} helped provide the necessary information to inform a followup decision on S250725j and S251112cm. Maintaining these communications with \ac{LVK} will be critical as new interferometer sensitivities increase the rate of detected \ac{GW} events during the LSST.

If possible, formal communications channels with the IceCube and \ac{Super-K} collaborations would benefit Rubin \ac{ToO} from a decision making perspective, to certify private information of specific alert qualities to best inform followup strategy. A significant portion of this expertise will come from the \ac{ToO} advisory committee, however, a point-of-contact looking at the data from the external experiment can provide valuable insight that should be sustained throughout the duration of the \ac{LSST}.

\item \textbf{Clear Communication and Engagement with the External Science Community}

Transparent communication channels with the external community foster trust and clarify operational performance for specific \ac{ToO}s. While much of the focus during the SV period was on delivering a working \ac{ToO} system, engagement with the community was not an emphasis. 

This should not be the status quo, and two methods of communication exist to work with the science community as a \ac{ToO} followup is happening:
\begin{itemize}
    \item \textbf{Rubin Community Forum:} When a {ToO} is triggered, a message on the Rubin Community Forum could notify the science community that a \ac{ToO} is being triggered, details on the intended observing strategy, and where to go to access relevant data products. This is the most straightforward route to build trust from the scientific community in Rubin \ac{ToO} responses.
    \item \textbf{The Rubin Schedule Viewer:} The Rubin observing schedule is publicly accessible through the \href{https://usdf-rsp.slac.stanford.edu/obsloctap/skymap}{\texttt{ObsLocTAP}} service.\cite{DMTN-263} This includes target names, from which \ac{ToO} observations can be inferred. While not a direct confirmation of a \ac{ToO} followup, \texttt{ObsLocTAP} provides a live view into the visit data for Rubin observatory observations, which can be cross-matched to recent \ac{ToO}'s of interest by members of the community through services such as The Gravitational Wave Treasure Map\cite{gwTreasure}. 
\end{itemize}

\item \textbf{Regular Mock Alert Exercises with External Partners}

Mock alert exercises are essential for validating end-to-end system behavior under realistic conditions. We learned the most from testing the system on-sky, but \ac{ToO} responses can be simulated without using valuable on-sky time. This is being explored for alert receipt and parsing in the Rubin \ac{ToO} mock-data-challenge \cite{RTN-107}. These exercises reveal integration gaps and uncover software inconsistencies.

Regular rehearsals ensure that both automated systems and human workflows remain operationally ready. In periods of low \ac{ToO} activity (such as when \ac{LVK} detectors are offline), a mock \ac{ToO} response should occur in the \ac{BTS} environment, including alert receipt, observing strategy generation, and execution of simulated observations. This controlled trial of the \ac{ToO} workflow will protect against bitrot, integrate new workflows and alert schemas, and validate observing strategies to drive the program forward as we begin the \ac{LSST}.

\item \textbf{Tolerance for Experimentation}

The \ac{LSST} \ac{ToO} program will benefit from a structured tolerance for experimentation. This tolerance was demonstrated through the SV period, enabling large improvements in our understanding of the system, and the first observations using Rubin Observatory with the express intent of searching for new discoveries in the universe. With an appetite for observing the exotic, this experimentation should be maintained by the \ac{ToO} advisory committee, and acted upon when deemed necessary. 

While the authority to intervene on non-standard targets is placed in the \ac{ToO} advisory committee, there must be a balance to still observe standard \ac{ToO}s and remain in the designated time allocation endorsed by the \ac{SCOC}. Maintaining this will be paramount to the success of the Rubin \ac{ToO} program, and achieving the core-science goals of the \ac{LSST}.

\end{enumerate}

\section{Conclusion}\label{sec:Conclusion}
We summarize the status of the Rubin Observatory \ac{ToO} system as follows:

\begin{enumerate}
    \item The foundation of the \ac{ToO} system is built on algorithms that can adjudicate follow-up decisions in a low-latency environment, while maintaining support of generic alerts manually submitted from serendipitous multi-messenger events.
    \item While many aspects of the \ac{ToO} system are automated, we retain human interaction through the \ac{ToO} advisory committee, which is empowered to manually activate the \ac{ToO} system and deploy custom observing strategies as needed. 
    \item The latency of external alert processing has been demonstrated to $<5$ seconds, and an evaluation of the total latency to begin on-sky observation is ongoing. Additional exercises, both legitimate and simulated, will add sufficient statistics to identify bottlenecks in specific processes.
    \item The unique science cases for \ac{ToO} responses have been integrated to Rubin observatory, and tested through both simulated responses, and three on-sky activations of the \ac{ToO} system.
    \item Rubin Observatory has already demonstrated the deepest wide area search for \ac{EM} counterparts of \ac{GW}s to date, including a search over $\sim850\deg^2$, in two Rubin bands to an average $5\sigma$ limiting magnitude of $m_g\sim24.46$ and $m_i\sim23.53$, in under three hours.
\end{enumerate}

Through the aforementioned commissioning activities and ongoing collaborations with external partners, the Rubin \ac{ToO} system is poised to lead multi-messenger discovery for the duration of the \ac{LSST}. Despite the demonstrated performance on-sky, additional development is ongoing to ensure that on-sky responses are efficient and accurate. This includes a simulated exercise conducted with the \ac{LVK} and IceCube collaborations. This exercise was executed during late 2025, prior to the S251112cm \ac{ToO} activation discussed in section \ref{subsec:S251112cm}. This exercise consisted of \ac{LVK} and IceCube sending simulated alerts to a Kafka stream over $\sim1$ month, and Rubin Observatory ingesting and processing these alerts in the \ac{BTS} environment.\cite{RTN-107} Analysis of the acquired alerts in ongoing, and includes an assessment of the sensitivity of the current \ac{ToO} strategies to proposed \ac{EM} emission models, a thorough accounting of latency to on-sky observation, and recommendations to maximize the first year of Rubin Observatory operations with the \ac{ToO} program. These results will be released publicly to enable broad community engagement and a deeper understanding of the Rubin Observatory \ac{ToO} system.

Significant opportunities for activating the \ac{ToO} system are on the horizon, as Rubin Observatory prepares to begin the \ac{LSST} in 2026. In late 2026, \ac{LVK} will begin an intermediate observing run, \ac{IR1}, providing an opportunity for Rubin Observatory to discover \ac{EM} counterparts to \ac{GW} candidates. Additionally, sustained operations of the IceCube Neutrino Observatory will bring forth the opportunity to identify the first \ac{EM} counterpart to a \ac{HEv} signal. Through these joint activities, and serendipity, Rubin Observatory is poised to lead multi-messenger astrophysics during the \ac{LSST} through effective and efficient deployment of the \ac{ToO} system.

\acknowledgments 
This material is based upon work supported in part by the National Science Foundation through Cooperative Agreements AST-1258333 and AST-2241526 and Cooperative Support Agreements AST-1202910 and AST2211468 managed by the Association of Universities for Research in Astronomy (AURA), and the Department of Energy under Contract No. DE-AC02-76SF00515 with the SLAC National Accelerator Laboratory managed by Stanford University. Additional Rubin Observatory funding comes from private donations, grants to universities, and in-kind support from LSST-DA Institutional Members. This work has been supported by the French National Institute of Nuclear and Particle Physics (IN2P3) through dedicated funding provided by the National Center for Scientific Research (CNRS). This work has been supported by STFC funding for UK participation in LSST, through grant ST/Y00292X/1.

Some of the results in this paper have been derived using the healpy and HEALPix package.

\bibliography{report} 
\bibliographystyle{spiebib} 

\end{document}